\documentclass[useAMS,usenatbib]{mn2e}
\usepackage{graphicx,times,array,ctable,amsmath,amssymb}
\usepackage{bbm}
\usepackage{url}

\newcommand{\Msun}{\mathrm{M}_{\odot}}

\newcommand{\Mpc}{\mathrm{Mpc}}

\newcommand{\dd}{\mathrm{d}}
\newcommand{\pd}{\partial}

\newcommand{\kappaCWL}{\kappa_{\mathrm{CWL}}}

\newcommand{\tr}{\mathrm{tr}}
\newcommand{\FOS}{\textsf{{\bf F}}_{\mathrm{OS}}}
\newcommand{\FLS}{\textsf{{\bf F}}_{\mathrm{LS}}}
\newcommand{\FOL}{\textsf{{\bf F}}_{\mathrm{OL}}}
\newcommand{\Feff}{\textsf{{\bf F}}_{\mathrm{eff}}}

\newcommand{\zL}{z_{\mathrm{L}}}
\newcommand{\zs}{z_{\mathrm{S}}}
\newcommand{\Atens}{\boldsymbol{\mathcal{A}}}

\newcommand{\PhiLOS}{\Phi_{\mathrm{LSS}}}
\newcommand{\kcutoff}{k_{\mathrm{cutoff}}}

\title[Effect of LSS on Cluster-Lensing Magnification]{The Effect of Large-Scale Structure on the Magnification of High-Redshift Sources by Cluster-Lenses}

\author[D'Aloisio, Natarajan \& Shapiro]{Anson
  D'Aloisio$^1$\thanks{Email: anson@astro.as.utexas.edu},  
	Priyamvada Natarajan$^{2,3}$, and Paul R. Shapiro$^1$\\
$^1$Department of Astronomy and Texas Cosmology Center, University of Texas, Austin, TX 78712, USA \\
$^2$Department of Physics, Yale University, PO Box 208120, New
    Haven, CT 06520-8120\\
$^3$Department of Astronomy, Yale University, PO Box 208101, New Haven, CT 06511}

\begin{document}

\maketitle

%The effects of this cosmic weak lensing (CWL) on the magnification of cluster-lenses can be quantified statistically in terms of the power spectrum of matter density fluctuations.  Using a simple model for the magnification profiles of cluster-lenses at $z=0.3-0.5$, we show that fluctuations in the magnification of sources at $z=6-10$ due to CWL are $\sim10-20(20-30)\%$ for typical absolute magnifications $\sim5(10)$

\begin{abstract}
Cluster gravitational lensing surveys like the \emph{Hubble Space Telescope (HST) Frontier Fields} survey will detect distant galaxies 10-50 times fainter than any yet discovered.  Using these surveys to measure the luminosity function of such faint, distant galaxies, however, requires that magnification maps built from the constraints of strongly-lensed images be accurate.  For models that assume the cluster and nearby (correlated) structures are the only significant sources of lensing, a potential source of error in these maps comes from the fact that light rays also suffer weak deflections by uncorrelated large-scale structure (LSS) along the line-of-sight, i.e. cosmic weak lensing (CWL).   To demonstrate the magnitude of this effect, we calculate the magnification change which results when the same cluster-lens is placed along different lines of sight.  Using a simple density profile for a cluster-lens at $z\sim0.3-0.5$ and the power spectrum of the matter density fluctuations responsible for CWL, we show that the typical magnifications of $\sim 5(10)$ of sources at $z=6-10$ can differ by $\sim10-20(20-30)\%$ from one line-of-sight to another.  However, these fluctuations rise to greater than order unity near critical curves, indicating that CWL tends to make its greatest contribution to the most magnified images.  We conclude that the neglect of CWL in determining the intrinsic luminosities of highly-magnified galaxies may introduce errors significant enough to warrant further effort to include this contribution in cluster-lens modeling.  We suggest that methods of modeling CWL in galaxy-strong-lensing systems should be generalized to cluster-lensing systems. 
\end{abstract}


\section{Introduction}
The search for the earliest galaxies has recently pushed the redshift horizon of direct, spectroscopically-confirmed detection to $z=7.51$, corresponding to $\sim700$ million years after the Big Bang \citep{2013Natur.502..524F}.  However, according to the theory of structure formation in the $\Lambda$CDM paradigm, the first galaxies formed even earlier and were typically less massive and less luminous than those already discovered.  These lower mass/lower luminosity galaxies, in fact, are predicted to have been in the majority, even at the highest redshifts already detected.  Indirect observational evidence for such a galaxy population below current detection limits may already exist as the sources of starlight necessary to finish reionizing the intergalactic medium (IGM) by $z \gtrsim 6$, while simultaneously accounting for the large electron scattering optical depth ($\tau_{\mathrm{es}}$) found by recent CMB polarization measurements \citep{2013ApJS..208...19H,2013arXiv1301.1228R}.  However, the theory of reionization on which that indirect evidence depends for its interpretation is, itself, uncertain because it requires knowing the efficiency of galactic halos as sources of ionizing starlight.  In that sense, extending the range of galaxy luminosity function observed at $z \gtrsim 6$ by direct detection will be an important check on the theory of reionization, and will lead to improvements in its predictions.

Quasar absorption spectra show that the IGM was almost completely reionized by $z\sim 6$ \citep[e.g.][]{2001AJ....122.2833F,2002AJ....123.1247F,2006AJ....132..117F,2001ApJ...560L...5D}, while recent CMB measurements of $\tau_{es}$ yield a reionization midpoint (redshift at which the IGM was half ionized) of $z\sim11$, based on simple parameterizations of the IGM ionized fraction \citep{2013ApJS..208...20B,2013arXiv1303.5076P}.  The sparsity of quasars observed above $z\sim6$ suggests that they probably did not produce enough ionizing photons to have reionized the IGM by themselves, so it is widely believed that the bulk of reionization was completed by ultraviolet (UV) radiation released into the IGM by star-forming galaxies at $z\gtrsim6$ \citep[e.g.][]{1987ApJ...321L.107S,1994ApJ...427...25S,1996ApJS..102..191G,2005MNRAS.356..596M,2009ApJ...692.1476C,2010AJ....139..906W,2012MNRAS.425.1413F}. 

In recent years, \emph{HST} imaging of the \emph{Hubble Ultra Deep Field} (\emph{HUDF}) has begun to probe the galaxy population at $z\sim 7-12$, spanning much of the redshift range in which reionization likely occurred \citep[see e.g.][]{2012arXiv1212.3869O,2013MNRAS.432.2696M,2013MNRAS.432.3520D,2013ApJ...768..196S,2013ApJ...763L...7E,2013arXiv1301.1228R}.  When combined with CMB constraints on $\tau_{es}$, these observations provide evidence for the unseen contribution of ionizing photons from star-forming galaxies with luminosities below \emph{HST} detection limits \citep{2013arXiv1301.1228R}.  In the near future, the much-anticipated \emph{JWST}\footnote{\url{http://www.jwst.nasa.gov/}} will make it possible to probe the galaxy population further down the luminosity function, closer to the dwarf-galaxies that may have dominated the ionizing photon budget during reionization.  

In the meantime, it is becoming clear that gravitational lensing by galaxy clusters offers a potentially powerful way to detect high-$z$ galaxies that are otherwise too intrinsically dim to be accessible to current telescopes.  If a galaxy is located along the line-of-sight behind a massive galaxy cluster, gravitational lensing can cause the image of the background galaxy to appear at multiple locations on the sky, with sizes that are either enlarged or reduced with respect to the image without lensing. For compact images, the factor by which a lensed image is enlarged or reduced is given by the local lensing magnification, $\mu$.  Since gravitational lensing preserves surface brightness, the enlarged images have observed (apparent) luminosities larger than their intrinsic luminosities by a factor of $\mu$, allowing cluster-lenses to function as a kind of natural telescope -- the so-called gravitational telescope.  The boost in flux from lensing magnification can also enable spectroscopic confirmation in cases where it would otherwise not be possible.     

Cluster-lenses have already been employed successfully to study lensed galaxies at $z\lesssim7$, and to find and characterize some of the most highly redshifted galaxy candidates detected to date\footnote{See e.g. the {\bf  C}luster {\bf L}ensing {\bf A}nd {\bf S}upernovae survey with {\bf H}ubble (\emph{CLASH}), which has as one of its main objectives the detection of galaxies at $z>7$ \citep{2012ApJS..199...25P}: \url{http://http://www.stsci.edu/~postman/CLASH/}.} \cite[e.g.][]{2006A&amp;A...456..861R,2007ApJ...663...10S,2008ApJ...678..647B,2008ApJ...685..705R,2009ApJ...690.1764B,2009ApJ...697.1907Z,
2010ApJ...720.1559B,2011MNRAS.414L..31R,2012ApJ...745..155H,2012ApJ...755L...7B,2012Natur.489..406Z,2012ApJ...745..155H,2012arXiv1211.2230B,2013ApJ...762...32C,2014ApJ...792...76B,2013A&amp;A...559L...9B}.  These applications of gravitational telescopes have raised the question: what if \emph{existing} facilities could be used to detect and study the faint galaxies which likely played a significant role in the EoR?  Motivated by this prospect, the Hubble Deep Fields Initiative (HDFI) Science Working Group unanimously recommended a plan for the \emph{Frontier Fields}\footnote{\url{http://www.stsci.edu/hst/campaigns/frontier-fields/}} that includes six \emph{HST} deep fields centered on cluster lenses, along with six parallel ``blank fields" (offset from the cluster lenses), to simultaneously exploit the Advanced Camera for Surveys (ACS) and Wide-Field Camera 3 (WFC3).  The cluster-lensing and blank field components of the program will offer complementary benefits.  The former could make possible the detection of high-$z$ galaxies with intrinsic luminosities 10-50 times fainter than any distant galaxy observed to date, while the latter will significantly improve statistics of high-$z$ galaxy samples by increasing the area of sky imaged at HUDF09 depths by a factor of three.     

In order to realize the full potential of cluster-lenses as gravitational telescopes, observers must be able to recover accurate intrinsic luminosities of magnified background galaxies.  In a given cluster-lensing system, intrinsic luminosities can only be recovered once a model of the mass distribution of the cluster and its corresponding lensing magnification map are constructed, allowing translation  between observed and intrinsic luminosities.  These models are built upon constraints provided by locations and redshifts of multiply-imaged background galaxies, but additional constraints such as weak-lensing shear measured at the outskirts of lenses, kinematics of cluster-member galaxies, and X-ray luminosities of intracluster gas, are also used when available (see recent review by \cite{2011A&ARv..19...47K} for more details). 

Intrinsic luminosities derived from cluster-lenses are subject to uncertainties associated with lens modeling.  A number of research groups have developed their own model-building methodologies that are currently being carefully compared.  As part of their recommendation report\footnote{\url{http://www.stsci.edu/hst/campaigns/frontier-fields/HDFI_SWGReport2012.pdf}}, the HDFI Science Working Group compared magnification maps constructed by three different research groups for two cluster-lenses: Abell 1689 and the Bullet Cluster.  Magnification maps of these clusters differ among the three modeling methodologies by $\sim 20-25\%$ for $\mu \lesssim5$,  fractional differences low enough to be comparable to expected photometry and distance modulus uncertainties\footnote{As mentioned in the report, we note that part of the discrepancy could arise from the various groups not using the same inputs for their modeling.}.  However, the differences increase with magnification, and can be as high as $\sim 40-60\%$ for $\mu \sim 30$.  While the bulk of background sources are typically magnified by factors of only 2-5 by a lensing cluster, the rare highest-$z$ objects that are of particular interest for EoR studies are expected to be boosted by factors of 10 or more.  Reducing the uncertainties associated with large magnification is clearly of great interest for the prospect of studying intrinsically faint galaxies.  Fortunately, deep \emph{HST} imaging will likely yield a large number of multiply-imaged background galaxies for each \emph{Frontier Fields} cluster, increasing the likelihood that uncertainties in their corresponding magnification maps can be further reduced through careful modeling.  In preparation for the \emph{Frontier Fields}, six modeling groups were selected to produce preliminary maps with existing \emph{HST} data for the \emph{Frontier Fields} clusters using the same set of input images, and a comparison project for reconstruction methodologies with simulated data is currently underway (Natarajan, P., Meneghetti, M., Coe, D., et al. 2014, in preparation).

%model builders almost always assume the cluster and its surrounding (correlated) structures to be the only significant sources of gravitational lensing.  However,

A substantial amount of work has been devoted towards developing cluster-lens modeling techniques, but a possible complication arises from the fact that, in addition to the dominant lensing effects of the cluster, light rays from background galaxies are subject to weak lensing by intervening LSS as they travel to the observer.  We shall henceforth refer to this weak lensing by LSS as cosmic weak lensing (CWL).  ``Parametric models" are those in which the mass distribution which contributes to the lens is modeled by a sum of analytic formulae (e.g. NFW profiles) for the cluster halo, the galaxies or groups within it, and sometimes even a merging cluster halo, in a single lens plane. Apart from the assumption of a single lens plane, this assumes that all contributions are associated with observed tracers of mass which are local to the cluster and, as such, cannot be said to model the CWL, which is fundamentally uncorrelated with the cluster-lens system.  On the other hand, even ``non-parametric" models, which divide the cluster-lens' mass or gravitational potential into a grid of pixels \citep[see e.g.][and references therein]{2011A&ARv..19...47K}, or into a sum of basis functions, still assume that the combination of cluster and CWL can be modeled accurately by a single lens-plane.  Both the lack of CWL component in parametric models and the single-lens-plane assumption implicit in non-parametric models could lead to significant errors in the inferred magnification maps.  Errors in magnification measurements could impact the determination of the high-$z$ luminosity function (and inferences about the EoR) not only through their impact on the inferred intrinsic luminosities of the sources, but also through the effective survey volume.  

The contribution of CWL to strong-lensing systems has been investigated previously in a variety of contexts \citep[see e.g.][]{1994ApJ...436..509S,1996ApJ...468...17B,1997ApJ...482..604K,2004PhRvD..70b3008D,2005ApJ...629..673M,2005ApJ...635L...1W,2007ApJ...654..714H,2007MNRAS.382..121H,2009MNRAS.398.1298P,Jullo2010Sci,2011MNRAS.411.1628D,2011ApJ...726...84W,2012MNRAS.421.2553X,2012MNRAS.420L..18H,2012MNRAS.424..325J,2014MNRAS.440..870T}.  However, the contribution of CWL to the \emph{magnification} of high-$z$ galaxies by cluster-lensing has not yet been quantified.  In a blank field, i.e. in the absence of cluster lensing, the statistical dispersion in magnification by CWL alone is expected to be as high as $\sim20\%$ for galaxies at $z=10$ \citep[e.g.][]{2011ApJ...742...15T}.  Thus, it is natural to ask whether CWL can contribute to the magnification of cluster-lensing systems at a similar level.  The answer to this question may help inform future efforts devoted to reducing uncertainties in lensing magnification maps.  

In this paper, we quantify the contribution of CWL to the magnification of cluster-lenses statistically in terms of the nonlinear power spectrum of matter density fluctuations. Specifically, we use the power spectrum to calculate the typical fluctuation (i.e. dispersion) in magnification which would be observed if the same cluster-lens were placed along different lines of sight.  As we will show, this fluctuation will depend on where the lensed galaxy is located on the sky relative to the cluster lens critical curves. 

The magnitude of this dispersion will indicate how important the CWL contribution is to the magnification maps of cluster-lensing systems.  We note that the dispersion we calculate in this way is not the same as the actual measurement error of the magnification caused by neglecting CWL.  Calculating the measurement error would require us create a mock data set based upon a known magnification map from a given realization of CWL and an assumed cluster density profile.  This mock data set could then be analyzed as if there is \emph{no} CWL effect, to determine the true measurement error introduced by the neglect of CWL.  That is beyond the scope of the present paper.  We will show, however, that the effect of CWL on the magnification is large enough to warrant future work along those lines.        

The remainder of this paper is organized as follows.  In \S \ref{SEC:Preliminaries}, we define some basic quantities in the formalism of gravitational lensing.  In \S \ref{SEC:Formalism}, we present our calculation of the magnification including strong lensing by a primary lens and weak lensing by intervening LSS, and derive relevant statistical quantities for our analysis.  In \S \ref{SEC:Results}, we present the main numerical results of this paper and in \S \ref{SEC:Conclusion} we close with a discussion on the implications of our results.        

For all numerical calculations we adopt a flat $\Lambda$CDM cosmology with parameters $\Omega_m=0.32$, $\Omega_\Lambda=0.68$, $H_0 = 100 h~\mathrm{km~s^{-1}~Mpc^{-1}}$, with $h=0.67$, $\sigma_8=0.83$ and $n_s=0.96$, consistent with the first cosmological results from the \emph{Planck} satellite mission \citep{2013arXiv1303.5076P}.  Throughout this paper, we work in units in which the speed of light is unity ($c=1$).  

\section{Preliminaries}
\label{SEC:Preliminaries}

\subsection{Lens equation and magnification in the thin-lens approximation}
\label{SEC:thinlens}

We begin by summarizing some standard definitions in the formalism of gravitational lensing by a single mass, e.g. a galaxy or cluster of galaxies.  The physical sizes of even the most massive galaxy clusters are always much smaller than the distance scales between observer, lens, and source.  In this regime, the gravitational deflection of a light ray by the massive object can be approximated as occurring instantaneously at a single redshift along the trajectory of the ray.  Under this so-called ``thin-lens approximation", the angular coordinates $\boldsymbol{\theta}$ of an observed image are related to the angular coordinates $\boldsymbol{\beta}$ of the source by the lens equation

\begin{equation}
\boldsymbol{\beta}=\boldsymbol{\theta}-\boldsymbol{\alpha}(\boldsymbol{\theta}),
\label{EQ:thinlenseq}
\end{equation}       
where $\boldsymbol{\alpha}(\boldsymbol{\theta})$ is the deflection angle, which can be written as $\alpha_i(\boldsymbol{\theta}) = \pd \psi(\boldsymbol{\theta}) / \pd \theta_i$, the gradient of the deflection potential, 

\begin{equation}
\psi(\boldsymbol{\theta}) = \frac{1}{\pi} \int \dd^2 \theta'~\kappa(\boldsymbol{\theta'}) \ln \left| \boldsymbol{\theta} - \boldsymbol{\theta'} \right|. 
\label{EQ:deflectionpotential}
\end{equation} 
Here, $\kappa(\boldsymbol{\theta}) \equiv \Sigma(\boldsymbol{\theta}) /\Sigma_{\mathrm{crit}}$ is the convergence -- the mass density of the lens projected along the line-of-sight, in units of the critical surface mass density, $\Sigma_{\mathrm{crit}}\equiv  D_{\mathrm{S}} / (4 \pi G~D_{\mathrm{L}} D_{\mathrm{LS}})$, where $G$ is Newton's gravitational constant, and $D_{\mathrm{S}}$, $D_{\mathrm{L}}$, and $D_{\mathrm{LS}}$ are the angular diameter distances between observer and source, observer and lens, and lens and source, respectively.  From equation (\ref{EQ:deflectionpotential}) it can be shown that $\psi$ and $\kappa$ satisfy the two-dimensional Poisson equation, $\pd^2 \psi(\boldsymbol{\theta}) / \pd \theta^2_1+\pd^2 \psi(\boldsymbol{\theta}) / \pd \theta^2_2  = 2 \kappa(\boldsymbol{\theta})$. 

An area element $\dd^2 \beta$ in the source coordinates is related to an area element $\dd^2\theta$ in the image coordinates by $\dd^2 \beta = \det(\boldsymbol{\mathcal{ A}})~\dd^2 \theta$, where $\boldsymbol{\mathcal{ A}} \equiv \dd \boldsymbol{\beta}/\dd \boldsymbol{\theta}$ is the Jacobian of the lens map, $\boldsymbol{\beta}(\boldsymbol{\theta})$.  Using the above definitions, the Jacobian in the thin-lens approximation can be written as

\begin{equation}
\Atens \equiv \begin{pmatrix} 1-\kappa + \gamma_1 & \gamma_2 \\ \gamma_2 & 1- \kappa-\gamma_1 \end{pmatrix}.
\end{equation}
Here, we have written $\Atens$ in terms of the two components of the complex valued shear $\gamma= \gamma_1 + i \gamma_2$, where $\gamma_1\equiv\left( \pd^2 \psi / \pd \theta^2_1 - \pd^2 \psi / \pd \theta^2_2 \right)/2$, and $\gamma_2\equiv \pd^2 \psi / \pd \theta_1 \pd \theta_2$.  The magnification, $\mu$, is defined as the ratio of image to source area elements,   

\begin{equation}
\mu = \frac{1}{\det(\boldsymbol{\mathcal{ A}}) }= \frac{1}{(1-\kappa)^2 - |\gamma|^2}. 
\end{equation}
Below, we will refer to the Jacobian $\boldsymbol{\mathcal{A}}$ as the inverse magnification tensor, since $\mu = \det(\boldsymbol{\mathcal{A}}^{-1})$.  In the case that $\mu$ does not vary significantly over the area of a source, i.e. for a ``compact" source, $\mu$ is approximately equal to the ratio of observed to intrinsic luminosities, $L_{\mathrm{obs}}/L \approx \mu$.  For extended sources, $L_{\mathrm{obs}}/L$ is instead written in terms of an integral over the surface brightness distribution of the source, $I^{(s)}(\boldsymbol{\beta})$, weighted by $\mu$: $L_{\mathrm{obs}}/L = \int \dd^2 \beta~I^{(s)}(\boldsymbol{\beta}) \mu(\boldsymbol{\beta})~/~\int \dd^2 \beta~I^{(s)}(\boldsymbol{\beta})$.   Finally, we note that the magnification can be either positive or negative. The sign of the magnification reflects the orientation of the image relative to the source.  We shall not be concerned with relative orientations here, so we will always consider the absolute value of the magnification.

\subsection{An illustrative lens model:  the NFW lens}
\label{SEC:CSLmodels}

In \S \ref{SEC:Results}, we employ a simple lens model to explore the contribution of CWL to cluster-lensing systems.  The model we use is derived from the \citet{1997ApJ...490..493N} (NFW) density profile for dark matter halos, in which the average mass density in a spherical shell of radius $R$ is parameterized by   

\begin{equation}
\rho( R )=\frac{\rho_s}{R/R_s \left(1+ R/R_s \right)^2}.
\end{equation}
The NFW profile is uniquely specified by the characteristic density and radius parameters, $\rho_s$ and $R_s$ respectively, but it is most often in the literature characterized by the concentration parameter\footnote{The spherical overdensity convention adopted here to define the radius of a halo is but one of several in the literature.  All of these conventions involve finding the radius at which the mean density in the sphere reaches some multiple of either the critical density or mean matter density of the universe.  Different conventions lead to different definitions of the concentration parameter, so some caution must be exercised when comparing to other works.}, $c_{200} \equiv R_{200}/R_s$, where $R_{200}$ is the radius at which the mean overdensity in a sphere centered on the halo reaches 200 times the critical density of the universe, $\rho_{\mathrm{crit}}(z)$.  In this convention, the mass of the halo is given by $M_{200}=200 \rho_{\mathrm{crit}}(z) 4 \pi R_{200}^3 / 3$.  According to simulations, typical concentration parameters for cluster halos with masses $M_{200}\sim 8\times10^{14}-2\times10^{15}\Msun$, and redshifts $z \sim 0.2-0.5$, range from $c_{200}=3-5$ [see e.g. \citet{2007MNRAS.381.1450N,2008MNRAS.387..536G,2008MNRAS.390L..64D,2012MNRAS.423.3018P}].         

\begin{figure}
\begin{center}
\resizebox{8.5cm}{!}{\includegraphics[angle=-90]{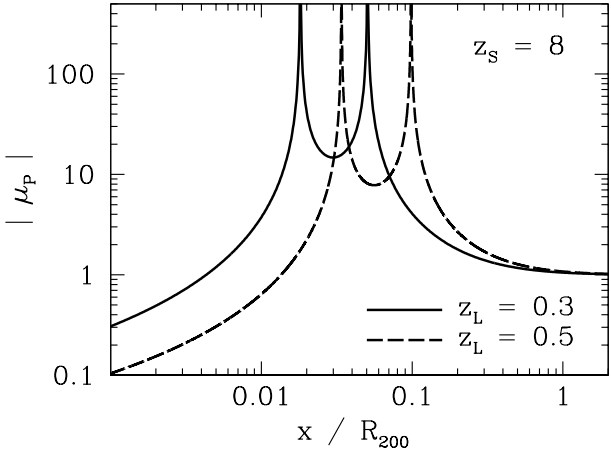}}
\end{center}
\caption{The lensing magnification profile of a dark matter halo whose mass density as a function of radius has the NFW form, with $M_{200}=2\times10^{15}\Msun$ and $c_{200}=4$ (the ``P" subscript in $\mu_P$ stands for ``primary lens", a distinction that will become important in the next section when we add LSS).  The impact parameter $x$ is given in units of $R_{200}$.  The solid and dashed curves correspond to lens redshifts of $z_L=0.3$ and $0.5$ respectively.  We assume a source redshift of $\zs=8$.  The peaks in the magnification profiles correspond to the critical curves of the lens.}
\label{FIG:magprofile}
 \end{figure}

The lensing characteristics, e.g. deflection angle, convergence, and shear, of the NFW mass profile were worked out in detail by \citet{1996A&A...313..697B} and \citet{2000ApJ...534...34W}.  In Figure \ref{FIG:magprofile}, we plot the absolute value of the magnification, $|\mu_P|$, of an NFW halo with mass $M_{200}=2\times10^{15}\Msun$ (typical of massive lensing clusters) and concentration parameter $c_{200}=4$, as a function of impact parameter $x$, in units of $R_{200}$.  Here, the ``P" subscript in $\mu_P$ stands for ``primary lens", a distinction that will become important in the next section when we add LSS.  We show results for two lens redshifts, $\zL=0.3$ (solid) and $\zL=0.5$ (dashed), assuming a source redshift of $\zs=8$.  For reference, $R_{200}=2.4(2.2)$ physical $\Mpc$, corresponding to angles of $\theta_{200}=8.6(5.8)'$ at $\zL=0.3(0.5)$.  

Note that the central regions of the lens (small values of $x/R_{200}$) correspond to magnifications less than unity.  Images located in these regions are de-magnified, so they appear smaller and dimmer with respect to the source.  The magnification formally diverges at locations where $\det(\boldsymbol{\mathcal{ A}})=0$.  The set of image coordinates at which this occurs constitute the critical curves of the lens, and the corresponding source coordinates are the caustics.  The sharp peaks in Figure \ref{FIG:magprofile} show that there are two circular critical curves in our illustrative lens model.  They occur at $\theta=9.4(12.9)''$ and $\theta=26.2(37.3)''$ for $\zL=0.3(0.5)$ and $\zs=8$.  The angle of the outer-most critical curve is called the Einstein radius of the lens, $\theta_{\mathrm{E}}= \left( 4 G M_{\mathrm{E}} D_{\mathrm{LS}}/D_{\mathrm{L}} D_{\mathrm{S}} \right)^{1/2}$, where $M_{\mathrm{E}}$ is the mass enclosed within $\theta_{\mathrm{E}}$. 

The caustics are demarcation points in the source plane for the formation of multiple images from a single source, with the outer-most caustic enclosing the area in which multiple images can be produced.   The image multiplicity is increased by 2 every time a test source crosses a caustic from the outside.  For the circularly symmetric models shown in Figure \ref{FIG:magprofile}, the inner-most caustic (the ``tangential caustic") is a single point at the center, so these models are capable of lensing a point source into 3 images (if the source does not sit exactly on any caustic).  However, when asymmetry is introduced to the model -- e.g. in the form of CWL -- the tangential caustic becomes an extended curve, allowing as many as 5 images to form, if the source sits within the area enclosed by the tangential caustic.  We note that $\kappa>1$ (i.e. surface mass density exceeds the critical value $\Sigma_{\mathrm{crit}}$) is in general a sufficient but not always necessary condition for the formation of multiple images. 

We conclude this section by noting that realistic parametric cluster-lens models are in fact much more complicated than the simple model considered here.  Out of necessity, realistic models include substructures associated with cluster galaxies \citep{1996ApJ...471..643K,1997MNRAS.287..833N}, and sometimes merging groups of galaxies, in order to reproduce observed image configurations.  Even the smoother cluster-halo component of the deflection potential typically requires a deviation from circular symmetry\footnote{To address this problem, \citet{2002A&A...390..821G} gave a useful prescription for adding ellipticity to the deflection potential of the NFW lens.}.  In this paper, our aim is to provide a simple assessment of CWL's contribution to the magnification of cluster-lensing systems.  The circularly symmetric NFW deflection potential suffices for this purpose. 

\section{Formalism}
\label{SEC:Formalism}

\subsection{Magnification of a thin lens including CWL}

It is necessary to extend equation (\ref{EQ:thinlenseq}) to include the effects of CWL with a primary thin-lens.  Our starting point is the lens equation of \citet{1996ApJ...468...17B}, which treats the primary lens under the thin-lens approximation, but also includes weak deflections accumulated along the trajectory of a light ray.  \citet{1996ApJ...468...17B} considered the impact of CWL on relative image positions and time delays in strong-lensing systems.  Here, we shall extend his calculation to quantify the effects of CWL on strong-lensing magnifications.  The lens equation of \citet{1996ApJ...468...17B} is given by  

\begin{align}
\boldsymbol{\beta}=  \boldsymbol{\theta} -\boldsymbol{\alpha}(\boldsymbol{\theta'}) + \FOS \cdot \boldsymbol{\theta} - \FLS \cdot \boldsymbol{\alpha}(\boldsymbol{\theta'}),
\label{EQ:Barkanalenseq}
\end{align}
where the deflection angle of the primary lens is evaluated at
\begin{equation}
\boldsymbol{\theta'} = \boldsymbol{\theta} + \FOL \cdot \boldsymbol{\theta},
\label{EQ:coord_thetaprime}
\end{equation}
Here, the effects of CWL are encapsulated in the tensors $\textsf{\bf F}_{\mathrm{AB}}^{i j} \equiv \textsf{\bf F}^{i j}{(\chi_\mathrm{A}, \chi_\mathrm{B}})$, where

\begin{align}
\textsf{\bf F}^{i j}({\chi_1, \chi_2}) \equiv - \frac{2}{\chi_2 -\chi_1} \int_{\chi_1}^{\chi_2} \dd \chi~ \left. \frac{\pd^2 \PhiLOS}{\pd x_i \pd x_j} \right|_{\boldsymbol{x}=\boldsymbol{0}}  \nonumber \\ \times (\chi_2 -\chi)( \chi-\chi_1),
\label{EQ:Ftensor}
\end{align}   
where $\chi(z)$ is the comoving distance between observer and redshift $z$, $\PhiLOS$ is the Newtonian potential from structures external to the primary lens plane\footnote{Note that $\PhiLOS$ satisfies the cosmological Poisson equation, $\nabla^2 \PhiLOS = 4 \pi G a^2 \bar{\rho} \delta$, where $a$ is the scale factor, $\bar{\rho}$ is the mean matter density, and $\delta=\rho/\bar{\rho} - 1$ is the density contrast.}, and $\boldsymbol{x}$ is the component of the comoving coordinate vector transverse to the line of sight, i.e. $\boldsymbol{x}=\boldsymbol{\theta} \chi_{\mathrm{L}}$, where $\chi_{\mathrm{L}}\equiv \chi(\zL)$ is the comoving distance between observer and the primary lens plane.  We note that the second derivative of $\PhiLOS$ which appears in the integrand of equation (\ref{EQ:Ftensor}) is evaluated at $\boldsymbol{x}=0$, since equations (\ref{EQ:Barkanalenseq}), (\ref{EQ:coord_thetaprime}), and (\ref{EQ:Ftensor}) were obtained by a Taylor expansion of $\PhiLOS$ about $\boldsymbol{x}=0$.        

%We use the shorthand notation $\textsf{\bf F}_{\mathrm{AB}}^{i j} \equiv \textsf{\bf F}^{i j}{(\chi_\mathrm{A}, \chi_\mathrm{B}})$ above and throughout the rest of this paper.   

The above lens equation was derived under the assumption that $\PhiLOS(\boldsymbol{x},\chi)$ is well-approximated by a linear expansion in the impact vector (i.e. transverse coordinates) $\boldsymbol{x}$.  This approximation is expected to break down at large angular separation from the lens center, where the variation of effective convergence/shear with angle becomes an important consideration.  \citet{1996ApJ...468...17B} showed how to avoid this assumption formally, but at the cost of adding a significant level of complexity to the equations.  In what follows, we will use equation (\ref{EQ:Barkanalenseq}) as a first approximation in the \emph{cores} of cluster-lenses, where images are typically located at angular radii of $\sim10-60''$ from the lens center (for a source at $\zs=8$, the Einstein radii of the lens models in \S \ref{SEC:CSLmodels} are $\theta_{\mathrm{E}} = 26.2''$ and $\theta_{\mathrm{E}}=37.3''$, for $\zL=0.3$ and $\zL=0.5$ respectively).  A calculation that is accurate at all angles from lens center requires numerical methods that are beyond the scope of this paper, e.g. ray tracing through cosmological structure formation simulations.  We note, however, that even a fully numerical approach would be subject to limitations from finite resolution of the simulation [see \citet{2011ApJ...742...15T} and \S \ref{SEC:smallscalestructure} of the current paper], as well as noise in ray-tracing calculations originating from discreteness of the density field in the N-body method \citep{2004A&amp;A...423..797B,2006ApJ...652...43L,2009MNRAS.398.1235X,2013MNRAS.430.2232R,2014MNRAS.444.2925A}.  Therefore, the semi-analytical results herein presented provide a useful first investigation, against which to compare future results from numerical methods.    

To compute the magnification, it is convenient to work in the image and source coordinates $\boldsymbol{\theta'}$ and $\boldsymbol{\beta'}$ respectively, where $\boldsymbol{\theta'}$ is given by equation (\ref{EQ:coord_thetaprime}), and

\begin{align}
\boldsymbol{\beta'} = \boldsymbol{\beta}-\FLS \cdot \boldsymbol{\beta}.
\label{EQ:coortrans}
\end{align}
In these coordinates, the lens equation takes the simplified form \citep{1996ApJ...468...17B}

\begin{equation}
\boldsymbol{\beta'}=\boldsymbol{\theta'} -  \boldsymbol{\alpha}(\boldsymbol{\theta'}) - \Feff \cdot \boldsymbol{\theta'},
\end{equation}
where $\Feff =-\FOS + \FLS + \FOL$, and the corresponding Jacobian is

 \begin{equation}
\frac{\dd \boldsymbol{\beta'}}{\dd \boldsymbol{\theta'}}=  \boldsymbol{\mathcal{ A}}_P.(\boldsymbol{\theta'}) - \Feff.
\label{EQ:transformedJac}
\end{equation}
Here, we have written the Jacobian in terms of $\boldsymbol{\mathcal{ A}}_P(\boldsymbol{\theta'})$, the inverse magnification tensor of the primary lens evaluated at $\boldsymbol{\theta'}$.  The coordinate transformation in equations (\ref{EQ:coord_thetaprime}) and (\ref{EQ:coortrans}) is useful because the magnification factor in the transformed coordinates, $\mu' \equiv 1/\det(\dd \boldsymbol{\beta'} / \dd \boldsymbol{\theta'})$, can be more easily computed than the observed magnification $\mu = 1/\det(\dd \boldsymbol{\beta} / \dd \boldsymbol{\theta})$.  The observed magnification then follows from the relation \citep{1997ApJ...482..604K}

\begin{equation}
\frac{1}{\mu (\boldsymbol{\theta})} = \frac{1}{\mu' (\boldsymbol{\theta'})} \frac{\det( \textsf{{\bf1}}+ \FOL)}{ \det(  \textsf{{\bf1}} - \FLS)}, 
\label{EQ:murelation}
\end{equation}
which can be shown using equations (\ref{EQ:coord_thetaprime}) and (\ref{EQ:coortrans}), and the multiplicativity of determinants.

Factoring out $\boldsymbol{\mathcal{ A}}_P(\boldsymbol{\theta'})$ on the left-hand side of equation (\ref{EQ:transformedJac}) and applying equation (\ref{EQ:murelation}), we find

\begin{align}
\frac{1}{\mu (\boldsymbol{\theta})} = \frac{1}{\mu_P (\boldsymbol{\theta'})} \det\left(  \textsf{{\bf1}} - \boldsymbol{\mathcal{ A}}_P^{-1}(\boldsymbol{\theta'}) \cdot \Feff \right)  \nonumber \\ \times   \frac{\det\left( \textsf{{\bf1}} + \FOL \right)}{\det\left( \textsf{{\bf1}} - \FLS \right)},
\end{align} 
where $\mu_P(\boldsymbol{\theta'})\equiv 1/\det(\boldsymbol{\mathcal{ A}}_P(\boldsymbol{\theta'}))$ is the magnification of the primary lens evaluated at $\boldsymbol{\theta'}$.  Finally, if we keep only terms that are linear in $\textsf{\bf F}^{ij}_{\mathrm{AB}}$ (which is justified by the fact that $\textsf{\bf F}^{ij}_{\mathrm{AB}}$ is typically less than 10 \% for the lensing configurations considered here -- see \S \ref{SEC:statistics}) , we are left with

\begin{align}
\frac{1}{\mu (\boldsymbol{\theta})} = \frac{1}{\mu_P (\boldsymbol{\theta'})} \biggl\{ 1 + \tr(\FOS)  + \tr\left[ \left(\textsf{{\bf 1}} - \boldsymbol{\mathcal{ A}}_P^{-1}(\boldsymbol{\theta'}) \right) \cdot \Feff \right] \biggr\},
\label{EQ:muinvResult}
\end{align}
where we have used the additivity of the trace and the definition, $\FLS+\FOL = \FOS + \Feff$.  

Equation (\ref{EQ:muinvResult}) can be evaluated in the case of a blank field, i.e. no primary lens, by setting $\kappa$ and $\gamma$ to zero, which implies $\boldsymbol{\mathcal{ A}}_P^{-1}=\textsf{{\bf 1}}$ and $\mu_P=1$, so $\mu = \left[ 1+\tr(\FOS) \right]^{-1}$, or 

\begin{equation}
\mu \approx 1 - \tr(\FOS).
\label{EQ:blankfieldMu}  
\end{equation}
This expression is equivalent to the standard expression for the lensing magnification to linear order in the convergence and shear (the so-called weak lensing limit), $\mu\approx 1+ 2 \kappaCWL$, where $\kappaCWL$ is the CWL effective convergence \citep[e.g.][]{2001PhR...340..291B}.  The connection is made explicit by inspection of equation (\ref{EQ:Ftensor}), and the definition

\begin{equation}
\kappaCWL =   \int_0^{\chi} \dd \chi' \left. \nabla^2_{\perp} \Phi \right|_{\boldsymbol{x}=\boldsymbol{0}} \left( 1 - \frac{\chi'}{\chi} \right) \chi',
\end{equation}   
where $\nabla^2_{\perp}=\pd^2/\pd x_1^2+\pd^2/\pd x_2^2$ .

\subsection{Statistics}
\label{SEC:statistics}

In the case of a blank field, we can use the standard deviation of $\mu$ as a measure of its statistical dispersion:

\begin{equation}
\sigma_{\mu}= \langle \left( \mu- 1 \right)^2 \rangle^{1/2}= \langle \tr\left( \FOS \right)^2 \rangle^{1/2},
\label{EQ:blankfield_stddev}
\end{equation}
where the average is over the ensemble of $\textsf{{\bf F}}^{ij}(\chi_1,\chi_2)$, and we have used the fact that $\langle \mu \rangle =1$ to the order of our calculation [see equation (\ref{EQ:blankfieldMu})].  We emphasize that our main focus is the impact of CWL on cluster-lensing systems, so our results include only fluctuations in $\mu$ that come from CWL.  We do not take into account additional strong-lensing by halos along the line-of-sight, but note that these events are rare anyway.  For example, \citet{2000ApJ...531..613B} calculate a strong-lensing optical depth of $\sim 1 \%$ for sources at $\zs \sim10$.  

In the case with cluster-lensing, we cannot easily Taylor expand equation (\ref{EQ:muinvResult}) and take expectation values to obtain the standard deviation of $\mu$ itself, since the components of $\boldsymbol{\mathcal{A}}_P^{-1}$ are proportional to $\mu$, and therefore diverge at the critical curves.  Instead, we quantify typical fluctuations in magnification maps with the fractional standard deviation (ratio of standard deviation to mean) of $\mu^{-1}$, 

\begin{equation}
\frac{\sigma_{(1/\mu)}}{| \langle \mu^{-1} \rangle |} = \frac{1}{| \langle \mu^{-1}(\boldsymbol{\theta}) \rangle |}\left\langle \left[\mu^{-1}(\boldsymbol{\theta})- \left\langle \mu^{-1}(\boldsymbol{\theta}) \right\rangle \right]^2 \right\rangle^{1/2}. 
\label{EQ:stddev}
\end{equation}
Before evaluating this expression with equation (\ref{EQ:muinvResult}), we examine the statistical properties of $\textsf{{\bf F}}^{ij}(\chi_1,\chi_2)$.  Doing so will allow us to make some simplying assumptions.   
  
 \begin{figure}
\begin{center}
\resizebox{8.5cm}{!}{\includegraphics[angle=-90]{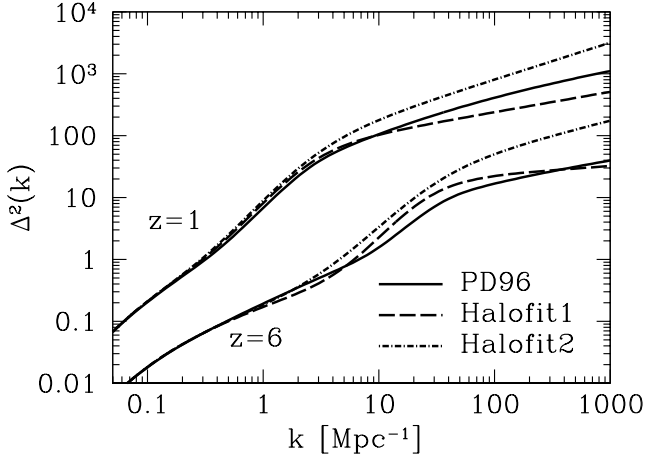}}
\end{center}
\caption{The power spectrum of matter density fluctuations at redshifts $z=1$ (top) and $z=6$ (bottom).  Here, we plot the dimensionless quantity, $\Delta^2(k)= k^3 P(k) / 2 \pi^2$, where $P(k)$ is the power spectrum.  At each redshift, we show three fits from dark-matter-only N-body simulations, which include the effects of nonlinear dynamics on the power spectrum.  The solid, dashed, and dot-dashed curves correspond to the fits of \citet{1996MNRAS.280L..19P} (PD96), \citet{2003MNRAS.341.1311S} (Halofit1), and \citet{2012ApJ...761..152T} (Halofit2), respectively.  For reference, at $z=1(6)$, $k=1~\Mpc^{-1}$ corresponds to an angular scale of $6.4(2.6)'$. }
\label{FIG:powerspectra}
 \end{figure}
 
 \begin{figure}
\begin{center}
\resizebox{8.5cm}{!}{\includegraphics[angle=-90]{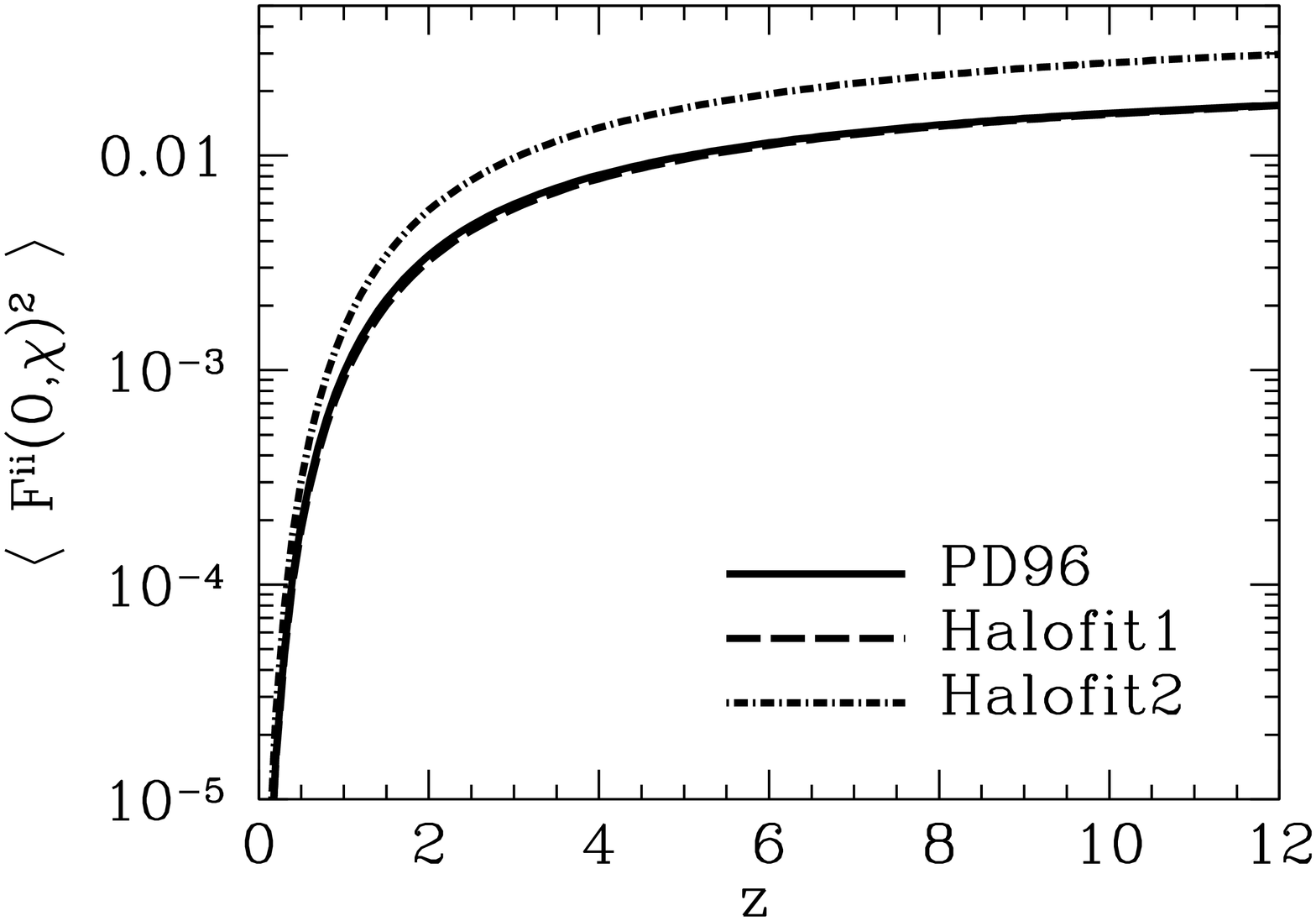}}
\resizebox{8.5cm}{!}{\includegraphics[angle=-90]{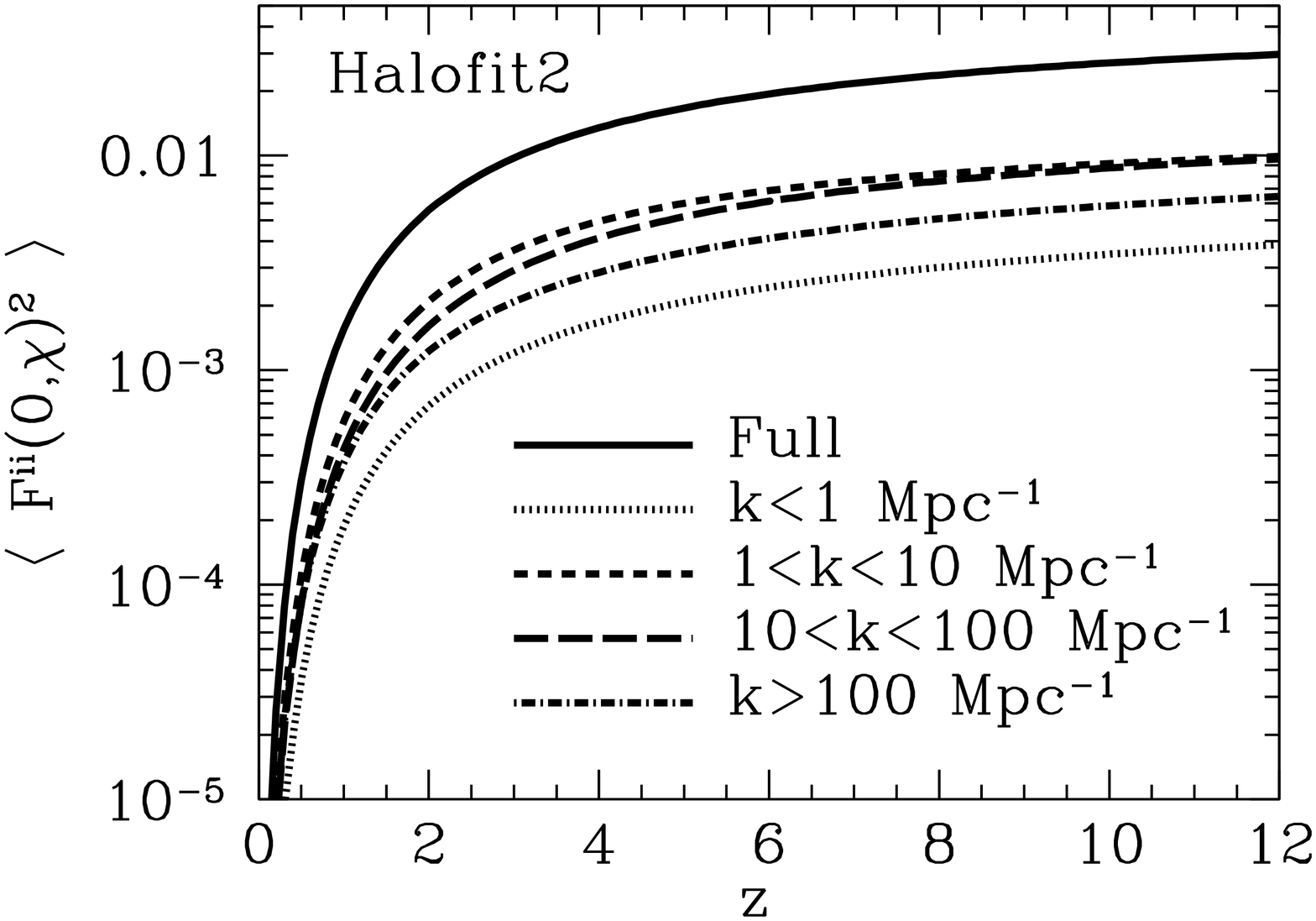}}
\resizebox{8.5cm}{!}{\includegraphics[angle=-90]{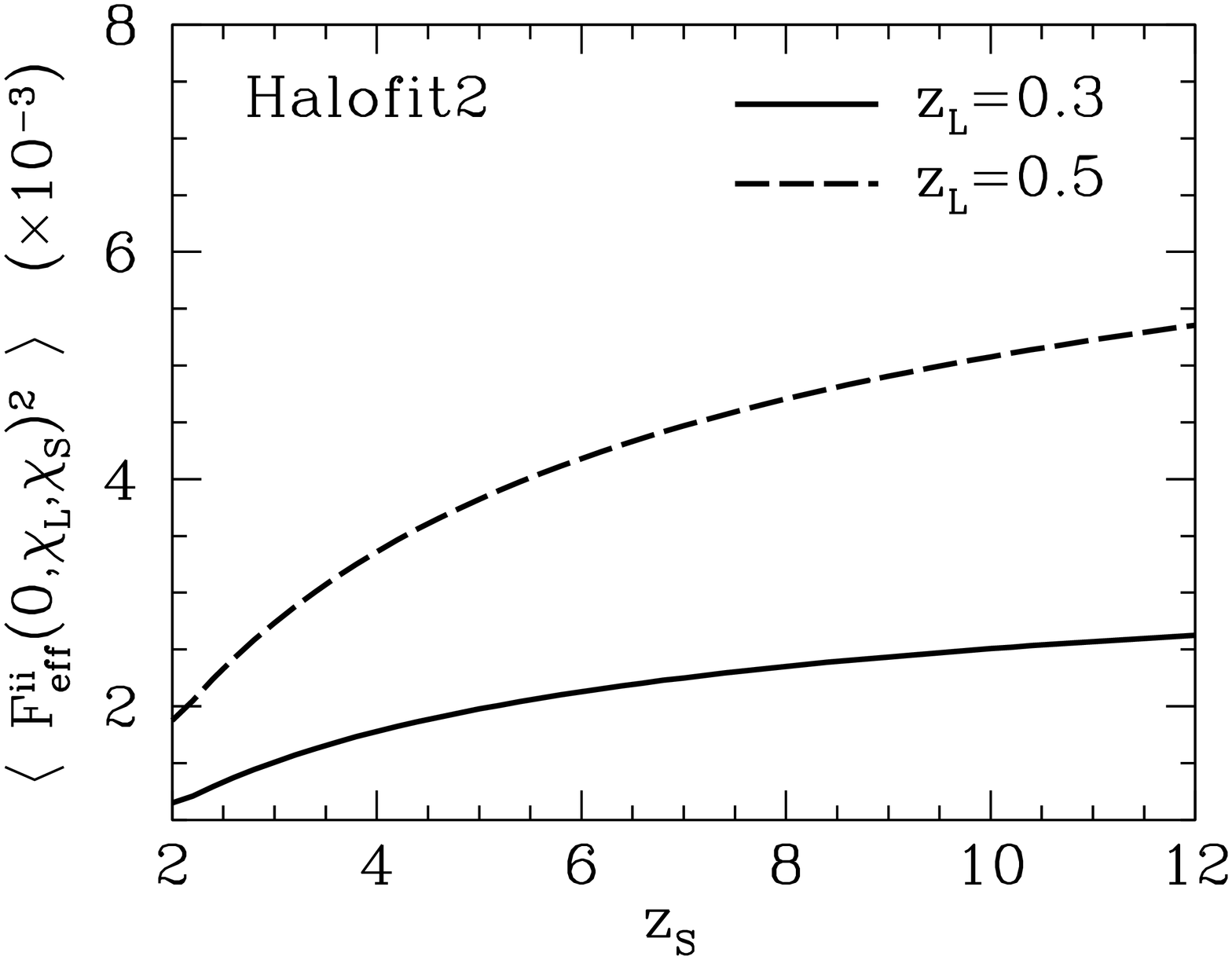}}
\end{center}
\caption{Variances of the LSS lensing tensors.  Top panel: variance of $F^{ii}(0,\chi)$ as a function of redshift for three different nonlinear power spectrum fits.  The variances of the diagonal components of $\FOL$ and $\FOS$ can be read from this plot.  For example, $\langle (\FOL^{11})^2 \rangle \approx 3\times10^{-4}$ at $\zL=0.5$, while $\langle (\FOS^{11})^2 \rangle \approx 2\times10^{-2}$ for $\zs=8$ (for Halofit2).  Middle panel:  contributions to $\langle F^{ii}(0,\chi)^2 \rangle$ from various $k$-ranges in the power spectrum.   Bottom panel:  variances of the effective convergence-shear tensor ($\Feff$) diagonal components as a function of $\zs$ for $\zL=0.3$ and $\zL=0.5$.  For clarity, we show only results from Halofit2 in the middle and bottom panels.}
\label{FIG:covariance}
 \end{figure}

There are three tensors in equation (\ref{EQ:muinvResult}) containing the effects of CWL:  $\FOS$ and $\Feff$, which appear explicitly, and $\FOL$, which appears implicitly through $\boldsymbol{\theta'}$ in the arguments of $\mu_P$ and $\boldsymbol{\mathcal{A}}_P^{-1}$.  The components of these tensors are random variables with zero mean, and their covariances can be written in terms of the power spectrum of the potential, $P_\Phi$, defined by $\langle \Phi(\boldsymbol{k})\Phi(\boldsymbol{k'}) \rangle = (2 \pi)^3 P_\Phi(k) \delta_{\mathrm{D}}(\boldsymbol{k}+\boldsymbol{k'})$, where $\delta_{\mathrm{D}}$ is the Dirac delta function \citep{1996ApJ...468...17B}. For example\footnote{Equation (\ref{EQ:covariance}) differs from equation (26) of \citet{1996ApJ...468...17B} in two ways: (i) Here we express the covariances in terms of comoving distances rather than conformal times. (ii) \citet{1996ApJ...468...17B} adopts the Fourier convention, $\Phi(\boldsymbol{r},\chi) = \int \dd^3 k \Phi(\boldsymbol{k},\chi) \exp\left( i \boldsymbol{k} \cdot \boldsymbol{r} \right)$, whereas we define $\Phi(\boldsymbol{r},\chi) = (2\pi)^{-3} \int \dd^3 k \Phi(\boldsymbol{k},\chi) \exp\left( i \boldsymbol{k} \cdot \boldsymbol{r} \right)$, so the normalizations of our equation (\ref{EQ:covariance}) and equation (26) of \citet{1996ApJ...468...17B} differ by a factor of $(2 \pi)^3$.}, if $\chi_1 \leq \chi_2 \leq \chi_3$, 

\begin{align}
\langle  \textsf{{\bf F}}^{ij}(\chi_1,\chi_2) \textsf{{\bf F}}^{kl}(\chi_1,\chi_3) \rangle = \frac{Q_{ijkl}}{4 \pi} \int_0^{\infty} \dd k~k^5 \int_{\chi_1}^{\chi_2} \dd \chi \nonumber \\ \frac{\left(\chi_2-\chi \right) \left( \chi-\chi_1 \right)}{\chi_2 - \chi_1} \frac{\left( \chi_3 -\chi \right) \left(\chi -\chi_1 \right)  }{\chi_3 - \chi_1} P_{\Phi}(k,\chi),
\label{EQ:covariance}
\end{align}   
where 
\begin{equation}
Q_{ijkl} =
\begin{cases}
3 & \text{if}~i,j,k,l~\text{are all equal}, \\
1&  \text{if, of}~i,j,k,l,~\text{two}=x~\text{and two}=y, \\
0 & \text{otherwise}. \\
\end{cases}
\label{EQ:Qfac}
\end{equation}
To the order of our calculation, we neglect all higher-order correlation functions appearing in equation (\ref{EQ:stddev}). 

As we illustrate numerically in the next section, the integral over $k$ in equation (\ref{EQ:covariance}) receives contribution from wavenumbers deep into the nonlinear regime, up to $k\sim1000~\Mpc^{-1}$.  Evaluating equation (\ref{EQ:covariance}) numerically therefore requires an accurate non-linear matter power spectrum for $0\leq z \leq \zs $ (where $\zs \sim 10$ in the cases of interest here), down to scales corresponding to $k\sim1000~\Mpc^{-1}$.  Unfortunately, the baryonic physics which significantly impacts the power spectrum at such large wavenumbers is still poorly understood, with little consensus among simulations performed so far.  As a result, no existing analytical fit to the power spectrum includes the effects of baryonic physics.  In what follows, we use three fits that are calibrated against dark-matter-only N-body simulations, and gauge the impact of uncertainties in the power spectrum at small scales by comparing results among the three fits: (i) \emph{PD96} -- an extension of the \citet{1991ApJ...374L...1H} scaling procedure by \citet{1996MNRAS.280L..19P} (ii) \emph{Halofit1} -- the ``Halofit" model of \citet{2003MNRAS.341.1311S} based on the halo model of dark matter clustering \citep{2000ApJ...543..503M,2000MNRAS.318..203S,2002PhR...372....1C} (iii) \emph{Halofit2} -- revision of the original Halofit1 parameters by \citet{2012ApJ...761..152T}.  

In Figure \ref{FIG:powerspectra} we plot the dimensionless matter power spectra, $\Delta^2 \equiv k^3 P(k) / 2 \pi^2$, in these models as a function of wavenumber.  We note that the smallest scales resolved in the simulations of \citet{2012ApJ...761..152T}, with the highest resolution among the above studies, correspond to $k\sim 40~\Mpc^{-1}$.  We are therefore extrapolating these models well beyond the $k$-range in which they are calibrated against simulations.  However, we emphasize that the point of considering these models is to explore how variation in small-scale power affects our main conclusions.  Moreover, as we will show in \S \ref{SEC:smallscalestructure}, most of the contribution to CWL magnification comes from $k \leq 40~\Mpc^{-1}$.      

In Figure \ref{FIG:covariance} we plot the variances of the $\FOL$, $\FOS$, and $\Feff$ diagonal components.  Note that these variances for $\FOL$ and $\FOS$ are of the form $\langle  \textsf{{\bf F}}^{ii}(0,\chi)^2 \rangle$, so the top panel of Figure \ref{FIG:covariance} shows results for both $\FOL$ and $\FOS$.  The variances of all off-diagonal components can be obtained by dividing the corresponding diagonal component result by a factor of 3 [see equation (\ref{EQ:Qfac})].  The top panel of Figure \ref{FIG:covariance} shows that the PD96 and Halofit1 power spectra yield very similar results.  This is due to a compensation effect.  While the Halofit1 spectrum tends to have more power relative to PD96 at intermediate scales, it has less power at the largest wave numbers.  On the other hand, Halofit2 always has significantly more power beyond $k\sim5~\Mpc^{-1}$ compared to both PD96 and Halofit1. This difference leads to a large boost in the variances.  For example, at $z=6(8)(10)$, $\langle  \textsf{{\bf F}}^{ii}(0,\chi)^2 \rangle$ evaluated with PD96 is larger than that of Halofit1 by only $2.6(1.7)(1.1)~\%$, whereas the Halofit2 result is larger than the Halofit1 result by $73(74)(74)~\%$.  In order to illustrate the relative contributions to $\langle  \textsf{{\bf F}}^{ii}(0,\chi)^2 \rangle$ of the power spectrum over various $k$-ranges, we integrate equation (\ref{EQ:covariance}) over the ranges: $k<1~\Mpc^{-1}$, $1<k<10~\Mpc^{-1}$, $10<k<100~\Mpc^{-1}$, and $k>100~\Mpc^{-1}$.  These ranges correspond to the dotted, short-dashed, long-dashed, and dot-dashed curves in the middle panel of Figure \ref{FIG:covariance}, respectively.  The solid curve shows the full result, i.e. integration over $k$ from 0 to $\infty$.  The bottom panel of Figure \ref{FIG:covariance} shows the variances of the diagonal components of $\Feff$ as a function of source redshift for two lens redshifts: $\zL=0.3$ (solid) and $\zL=0.5$ (dashed).  For clarity, we show only the results from Halofit2 in the middle and bottom panels. 

Let us now return to the question of how to evaluate equation (\ref{EQ:stddev}) with equation (\ref{EQ:muinvResult}).  Taking the average of $\mu^{-1}(\boldsymbol{\theta})$ with equation (\ref{EQ:muinvResult}) is not straightforward due to its implicit dependence on $\FOL$.  Fortunately, Figure \ref{FIG:covariance} informs us how to simplify $\mu^{-1}(\boldsymbol{\theta})$ for our particular application.  Figure \ref{FIG:covariance} shows that the variances of the components of $\FOL$ are always much smaller than those of $\FOS$ and $\Feff$ for lensing configurations in which $\zL=0.3-0.5$, and $\zs \sim 6-10$, the lens and source redshift ranges relevant for the \emph{Frontier Fields}, for example.  Moreover, for those redshift ranges, we find that $| \langle \FOL^{ij} \FOS^{kl} \rangle|$ and $|\langle \FOL^{ij} \Feff^{kl}\rangle|$ are at least a factor of 10 smaller than $|\langle \FOS^{ij} \Feff^{kl} \rangle |$.  For example, using Halofit2 with $\zL=0.5$ and $\zs=6$, $\langle \FOL^{11} \FOS^{11} \rangle = 6.5\times10^{-4}$ and $\langle \FOL^{11} \Feff^{11}\rangle = -3.4\times10^{-4}$, whereas $\langle \FOS^{11} \Feff^{11}\rangle = -7.6\times10^{-3} $.  We therefore proceed by setting $\mu_P(\boldsymbol{\theta'})\approx \mu_P(\boldsymbol{\theta})$ and $\boldsymbol{\mathcal{ A}}_P^{-1}(\boldsymbol{\theta'})\approx \boldsymbol{\mathcal{ A}}_P^{-1}(\boldsymbol{\theta})$ in equation (\ref{EQ:muinvResult}), keeping in mind that this approximation breaks down for lensing configurations with higher-redshift lenses and/or lower-redshift sources.  With these approximations, $\langle \mu^{-1}(\boldsymbol{\theta})\rangle = \mu_P^{-1}(\boldsymbol{\theta})$, and equation (\ref{EQ:stddev}) yields

\begin{align}
\frac{\sigma_{(1/\mu)}}{|\langle \mu^{-1} \rangle|} = \biggl\{ \langle \tr(\FOS)^2 \rangle + \langle\tr[(\textsf{{\bf1}}- \boldsymbol{\mathcal{ A}}_P^{-1}(\boldsymbol{\theta}))\cdot \Feff]^2 \rangle \nonumber \\ + 2 \langle \tr(\FOS)~\tr[(\textsf{{\bf1}}- \boldsymbol{\mathcal{ A}}_P^{-1}(\boldsymbol{\theta}))\cdot \Feff] \rangle \biggr\}^{1/2}.
\label{EQ:CSL_sigmaMu}
\end{align}
In a blank field, the second and third terms on the right-hand side of equation (\ref{EQ:CSL_sigmaMu}) vanish and $|\langle \mu^{-1} \rangle| = 1$, so equation (\ref{EQ:CSL_sigmaMu}) reduces to $\sigma_{(1/\mu)} = \langle \tr(\FOS)^2 \rangle^{1/2}$, implying that $\sigma_\mu = \sigma_{(1/\mu)}$ to linear order.

In the next section, we use the NFW model described in \S \ref{SEC:CSLmodels} to evaluate equation (\ref{EQ:CSL_sigmaMu}).  CWL spoils the circular symmetry of the magnification in the image plane that is normally exhibited by the NFW lens, so the dispersion of $1/\mu$ in general depends on the polar angle of the image coordinates.  Since our main goals here are: (i) to provide a simple illustration of CWL's effect on cluster-lensing magnification, and (ii) to stimulate future numerical investigation using more realistic cluster-lens models, which are asymmetric even without the effects of CWL, we will numerically evaluate equation (\ref{EQ:CSL_sigmaMu}) along the $\theta_{1}$-axis.  Doing so simplifies our analysis considerably, because the circular symmetry of the NFW model renders $\boldsymbol{\mathcal{ A}}_P^{-1}$ diagonal on the $\theta_1$ axis, i.e. $\boldsymbol{\mathcal{ A}}_P^{-1}(\theta,0)= \mathrm{diag}\left[ \mu(1-\langle \kappa \rangle),\mu(1+\langle \kappa \rangle - 2 \kappa) \right]$, where $\langle \kappa \rangle$ is the mean convergence inside a circle of radius $\theta$.

\section{Results}
\label{SEC:Results}

\begin{figure}
\begin{center}
\resizebox{8.5cm}{!}{\includegraphics[angle=-90]{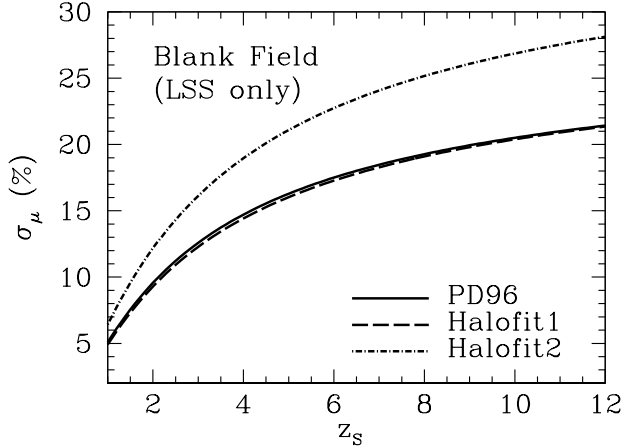}} 
\end{center}
\caption{Fractional standard deviation of the CWL magnification in a blank field (no cluster lens) as a function of source redshift.  The observed luminosity of a galaxy with intrinsic luminosity $L$ is $L_{\mathrm{obs}} = \mu L$.  So, for example, fluctuations in the lensing magnification of $\sim20-25\%$ for $\zs=10$ translates to a $\sim20-25\%$ uncertainty in the intrinsic luminosity of a source at that redshift.}
\label{FIG:sigmaMUblankfield}
\end{figure}
  
 \subsection{Blank field}
Before presenting our main results, it is instructive to consider the case of a blank field.  In Figure \ref{FIG:sigmaMUblankfield}, we evaluate equation (\ref{EQ:blankfield_stddev}) for the fractional standard deviation of the CWL magnification as a function of source redshift.  The results obtained from the PD96 and Halofit1 power spectra are very similar due to the compensation effects discussed in the last section, while the Halofit2 results are significantly higher due to the larger amount of small-scale power in that model relative to the others (see Figs. \ref{FIG:powerspectra} and \ref{FIG:covariance}). These differences highlight the level to which the results presented here and in the next section are sensitive to the effects of nonlinearities and baryonic physics on the power spectrum at large wavenumber.  Bearing these uncertainties in mind, we quote the range of results spanned by PD96, Halofit1, and Halofit2.   

The fractional standard deviation of $\mu$ rises from $9-12~\%$ for sources at $\zs \sim 2$, up to $\sim 20-30~\%$ for $\zs \gtrsim 10$.  Since the observed luminosity of a source with intrinsic luminosity $L$ is given by $L_{\mathrm{obs}}=\mu L$, the dispersion in $\mu$ results in an irreducible uncertainty in $L$.  So, for example, fluctuations in the CWL magnification of $\sim20-25\%$ for $\zs=10$ translates to a $\sim 20-25\%$ uncertainty in the intrinsic luminosity of a source at that redshift.  Note that this kind of uncertainty in luminosity measurements has been studied in detail at lower redshift in the context of surveys which aim to use Type IA Supernovae as standard candles \citep[e.g.][]{1996ComAp..18..323F,1997ApJ...475L..81W,1998ApJ...506L...1H,1999MNRAS.305..746M,2003ApJ...585L..11D}.  For similar cosmological parameters, we find that our results are consistent with those previous studies at lower redshift, and the more recent numerical work of \citet{2011ApJ...742...15T} over the full redshift range in Figure \ref{FIG:sigmaMUblankfield}.

\subsection{Cluster strong lensing}

\begin{figure}
\begin{center}
\resizebox{8.5cm}{!}{\includegraphics{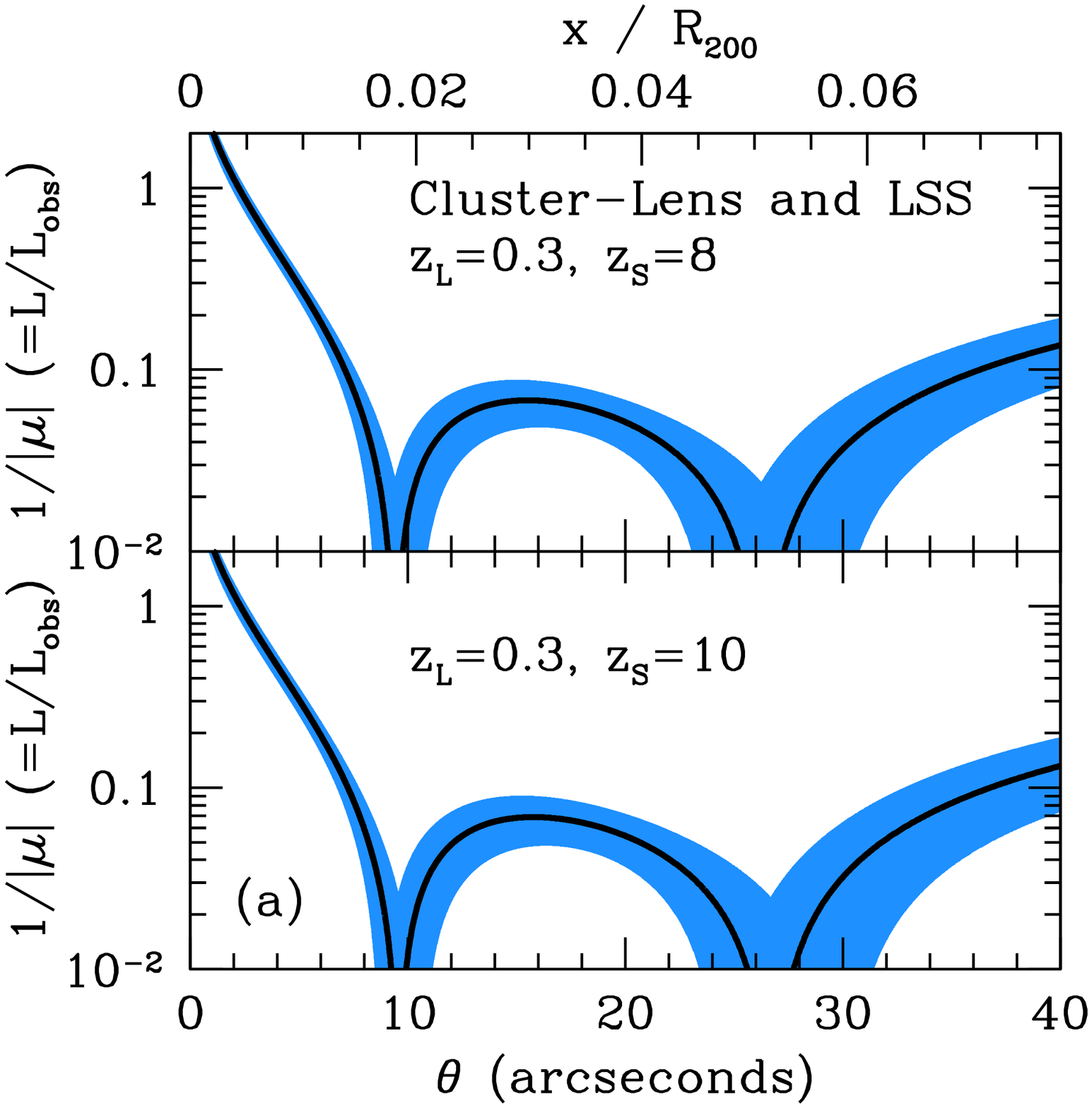}}\vspace{0.2cm}
\resizebox{8.5cm}{!}{\includegraphics{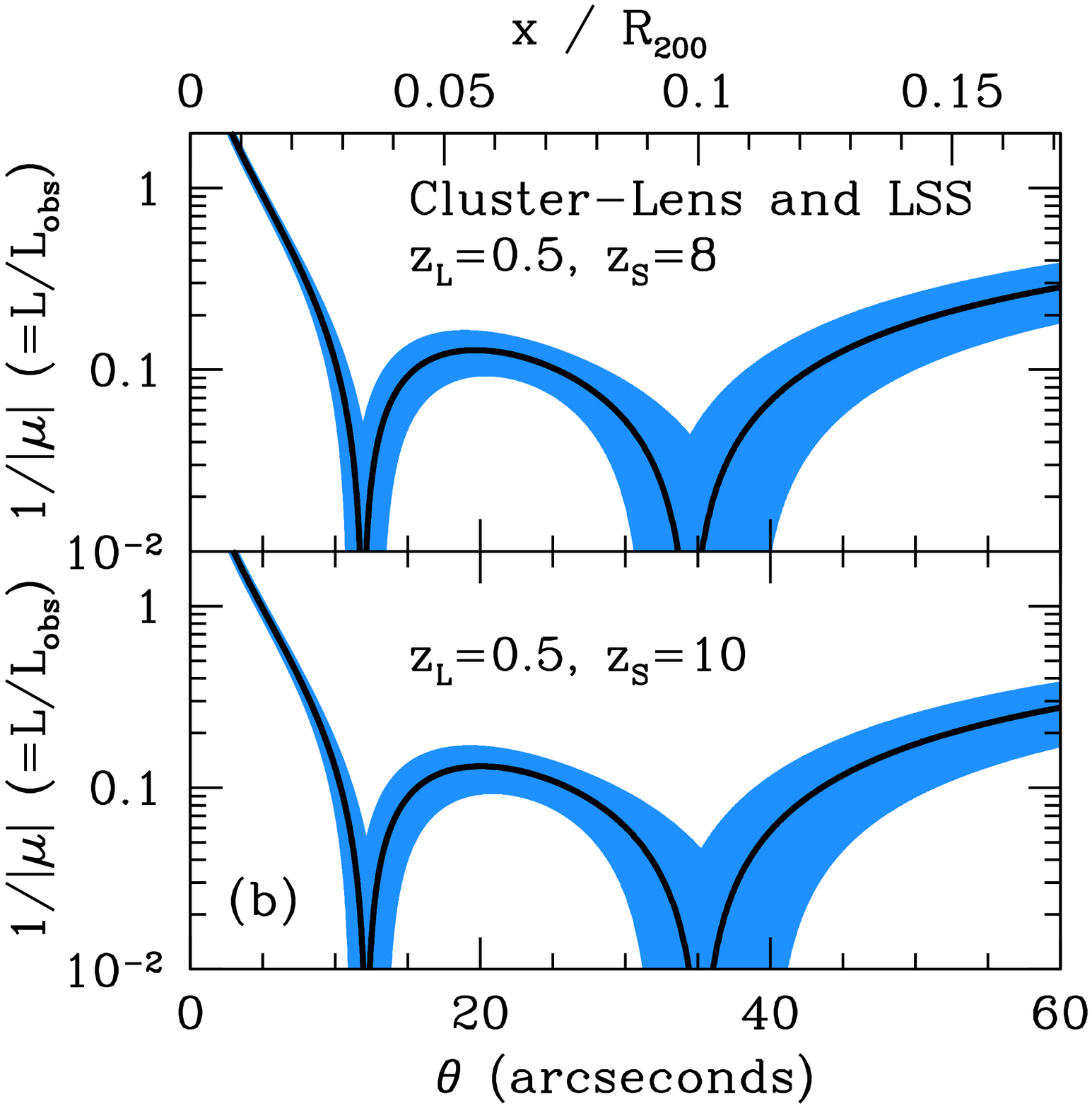}}
\end{center}
\caption{The effect of LSS on the magnification of high-$z$ sources by cluster-lenses I.  The ratio of intrinsic to observed luminosities for compact images, $L/L_{\mathrm{obs}}=|\mu^{-1}|$.  The solid lines show $|\mu^{-1}_P|$, the inverse magnification of the cluster-lens without the effect of LSS included, as a function of angular displacement from lens center, $\theta$.  For the cluster mass-profile, we employ the NFW model described in \S \ref{SEC:CSLmodels}, with $M_{200}=2\times10^{15}\Msun$ and $c_{200}=4$.   The shaded bands correspond to $<1\sigma$ fluctuations in $\mu^{-1}$ from CWL, calculated from the Halofit2 model of the matter power spectrum.  }
\label{FIG:flucbands}
 \end{figure}

\begin{figure}
\begin{center}
\resizebox{8.5cm}{!}{\includegraphics{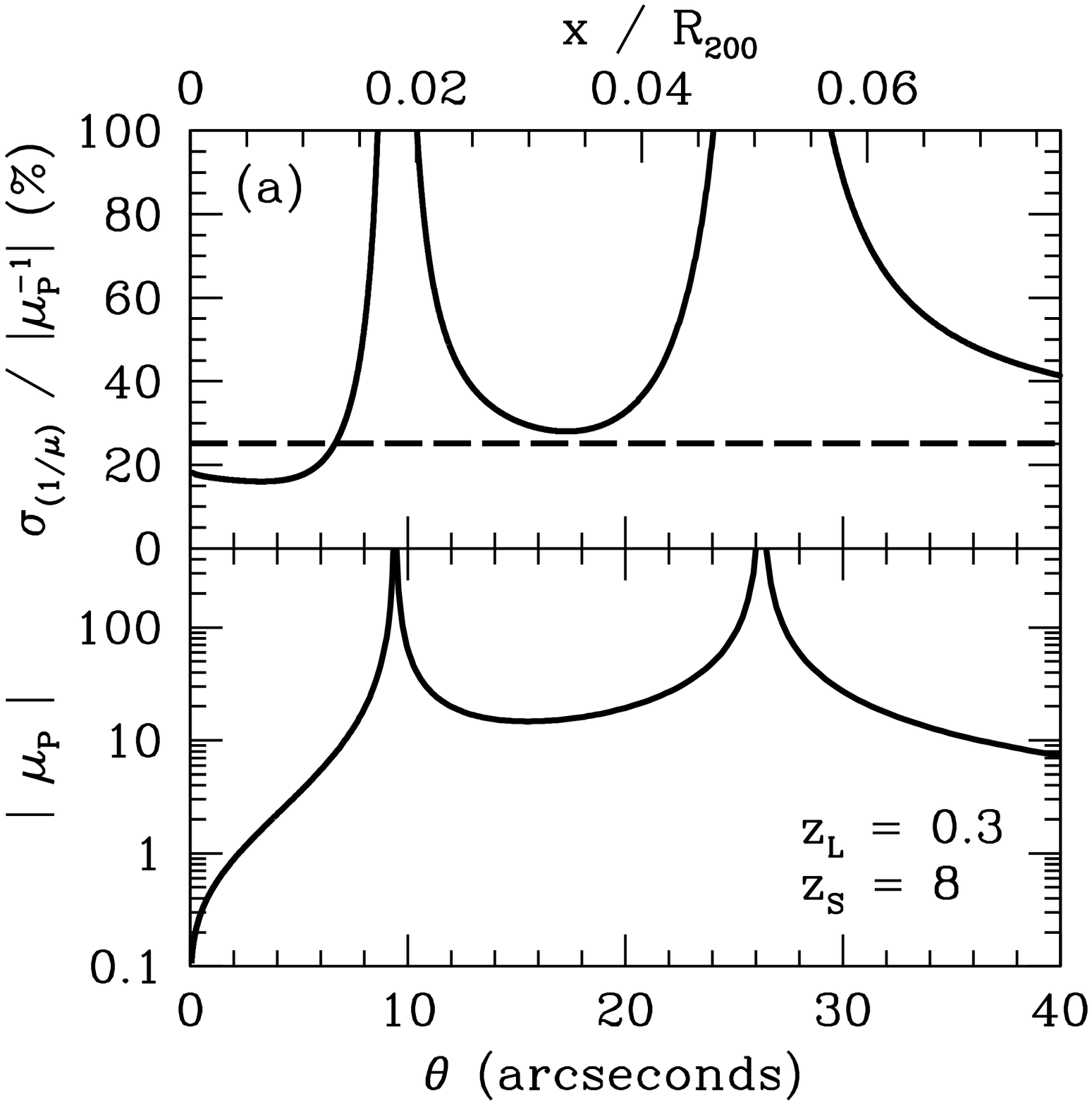}}\vspace{0.2cm}
\resizebox{8.5cm}{!}{\includegraphics{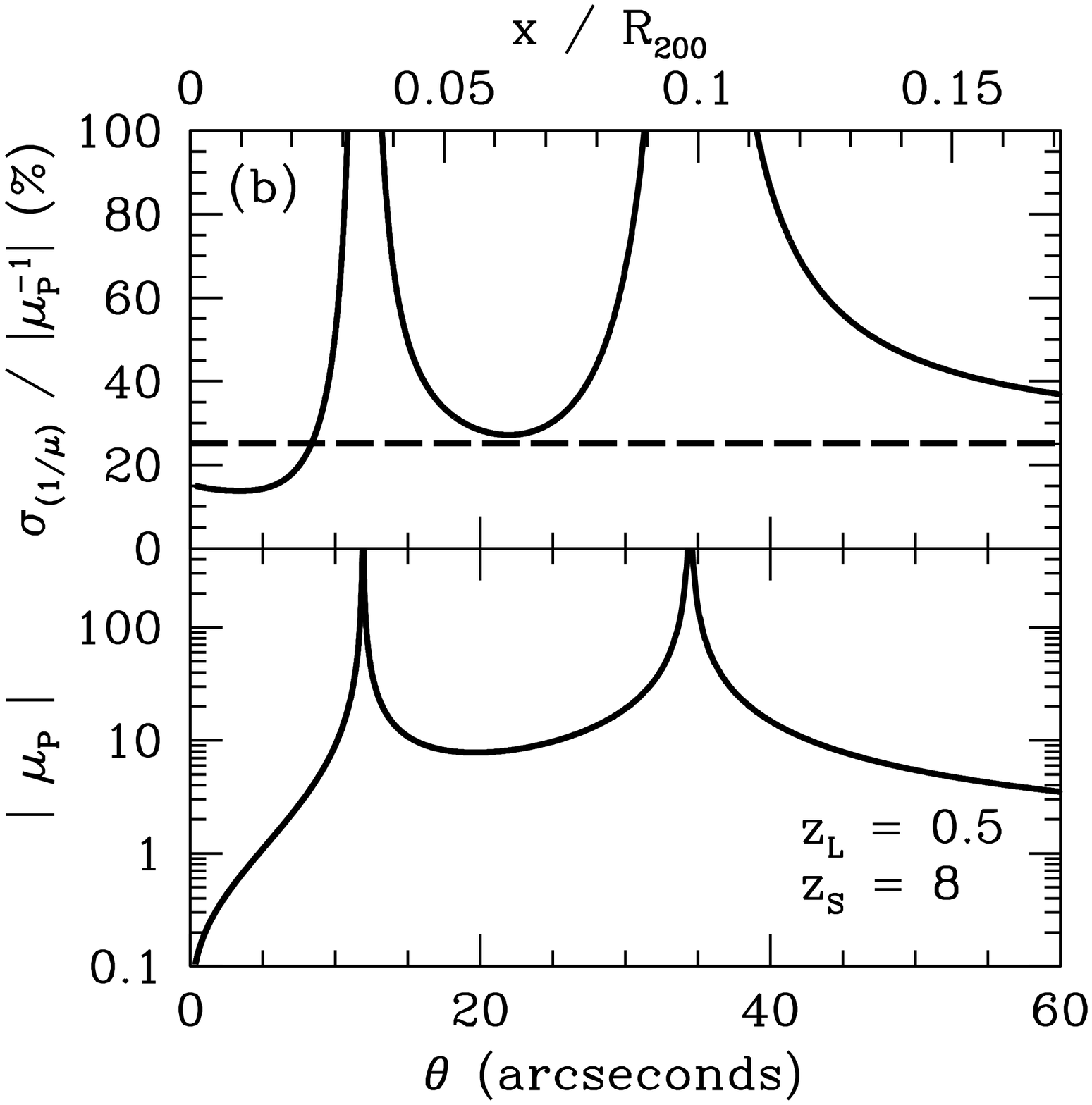}}
\end{center}
\caption{The effect of LSS on the magnification of high-$z$ sources by cluster-lenses II.  Top halves of (a) and (b): fractional standard deviation of $\mu^{-1}$ (ratio of the standard deviation of $\mu^{-1}$ to the mean, $|\langle\mu^{-1} \rangle|= |\mu_P^{-1}|$), as a function of angular displacement from lens center, in an illustrative cluster-lensing model with $\zL=0.3$ (a) and $\zL=0.5$ (b), and $\zs=8$.  The dashed lines correspond to the blank field results (see Figure \ref{FIG:sigmaMUblankfield}).  In the bottom halves of (a) and (b), we show the magnification profiles of the cluster-lenses without the effect of LSS included.}
\label{FIG:CSL_sigmaMuInv}
 \end{figure}

For the case with cluster-lensing, we employ the NFW model with $M_{200}=2\times10^{15}\Msun$ and $c_{200}=4$.  Figure \ref{FIG:flucbands} illustrates the statistical effect of CWL on magnification profiles of cluster-lenses.  There, we plot the ratio of intrinsic to observed luminosities for compact images, i.e. the inverse of the magnification, $|\mu|^{-1}$.  The solid lines show $|\mu^{-1}_P|$, the inverse magnification of the cluster-lens without the effects of CWL, as a function of angular displacement from lens center, $\theta$.  The shaded bands correspond to $<1\sigma$ fluctuations in $\mu^{-1}$ from CWL, calculated from equation (\ref{EQ:CSL_sigmaMu}) evaluated along the $\theta_{1}$-axis (see last paragraph of \S \ref{SEC:statistics}).  We show only results from Halofit2 for clarity.  Figure \ref{FIG:CSL_sigmaMuInv} provides a more quantitative representation of the fluctuations from CWL.  The top halves of Figures \ref{FIG:CSL_sigmaMuInv}a and \ref{FIG:CSL_sigmaMuInv}b show the fractional standard deviation of $\mu^{-1}$  as a function of $\theta$, for sources at $\zs=8$.  For reference, the dashed lines show the corresponding $\sigma_{(1/\mu)}$ for a blank field.  Again, we show only results from Halofit2 for clarity, but the corresponding results for PD96 and Halofit1 can be estimated by moving the dashed lines downwards according to the blank field results in Figure \ref{FIG:sigmaMUblankfield}.  The bottom halves of Figure \ref{FIG:CSL_sigmaMuInv}a and \ref{FIG:CSL_sigmaMuInv}b show the magnification profiles of the cluster-lenses without the effects of CWL.   

The standard deviation of $\mu^{-1}$ measures the level of typical fluctuations from CWL in the magnification profiles of the cluster-lenses.  According to Figures \ref{FIG:flucbands} and \ref{FIG:CSL_sigmaMuInv}, these fluctuations tend to be lowest in the central-most regions of the image plane, where images are demagnified and their detection is often made difficult by the presence of bright cluster-galaxies.  For lenses at $\zL\sim 0.3-0.5$ and sources at $\zs \sim6-10$, the fluctuations from CWL are $\sim10-20(20-30)\%$ for more typical magnifications of $|\mu|\sim 5(10)$, considering the range spanned by the PD96, Halofit1, and Halofit2 models.  However, the impact of CWL is greatest near the critical curves, as indicated by the steep rise in fluctuations approaching those locations.  This phenomenon results from the fact that CWL can perturb critical curves.  If an image-plane coordinate is near what would have been a critical curve in the absence of CWL (i.e. for the cluster in isolation, without intervening LSS), then the addition of CWL could move the critical curve closer to, further away from, or even on top of that coordinate.  This effective variation in the distance to the critical curve results in a large dispersion that increases as the critical curve is approached, since the magnification rises steeply in its vicinity.   

\begin{figure}
\begin{center}
\resizebox{8.5cm}{!}{\includegraphics{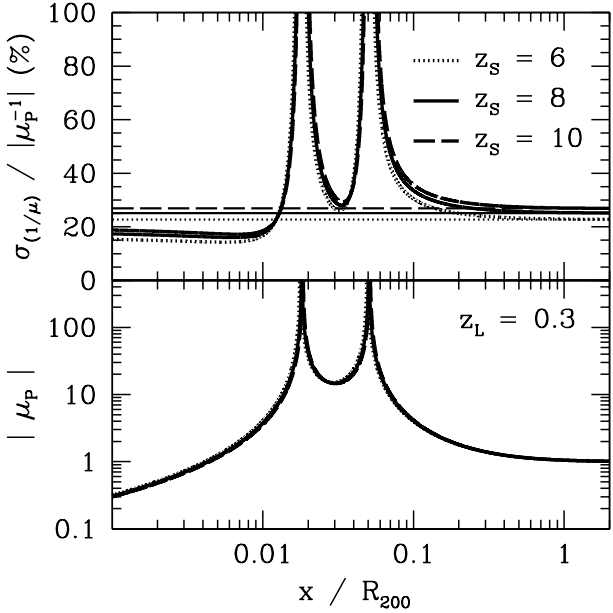}}
\end{center}
\caption{Same as Figure \ref{FIG:CSL_sigmaMuInv}, but plotted as a function of impact parameter $x$ (in units of $R_{200}$) for a fixed $\zL=0.3$, and $\zs=6$, 8, and 10. The horizontal lines show the corresponding blank field results.}
\label{FIG:CSL_sigmaMuInv_zs}
 \end{figure}

The blank field $\sigma_{(1/\mu)}$ sets the overall normalization of the dispersion profiles.  In fact, the fractional standard deviation converges to the blank field result at large radii from the lens center (see Figure \ref{FIG:CSL_sigmaMuInv_zs}).  This expected behavior at large radii appears to be qualitatively correct, but we caution that the accuracies of our calculations are expected to degrade with increasing radii from the lens center (see discussion in \S \ref{SEC:Formalism}).  Finally, we note that these results are only modestly dependent on the source redshift for $z_S\gtrsim6$, as demonstrated in Figure \ref{FIG:CSL_sigmaMuInv_zs}, where we plot the fractional standard deviations for a fixed $\zL=0.3$, while varying the source redshift between $\zs=6$, $8$, and $10$.  This modest dependence on $\zs$ is the result of two effects:  (i) As shown in the bottom panel of Figure \ref{FIG:CSL_sigmaMuInv_zs}, the magnification profiles of the primary lens do not vary much between $\zs=6$, $8$, and $10$, since the ratio of angular diameter distances appearing in the lens equation, $D_{\mathrm{LS}}/D_{\mathrm{S}}$, grows very slowly after $\zs=6$.  (ii) $\sigma_{(1/\mu)}$ in a blank field, which sets the overall normalization of the dispersion profile, rises slowly for $\zs>6$ (see Figure \ref{FIG:sigmaMUblankfield}).   

\subsection{The contribution of small-scale structure to CWL magnification}  
\label{SEC:smallscalestructure}

\begin{figure}
\begin{center}
\vspace{-0.5cm}
\resizebox{8.5cm}{!}{\includegraphics[angle=-90]{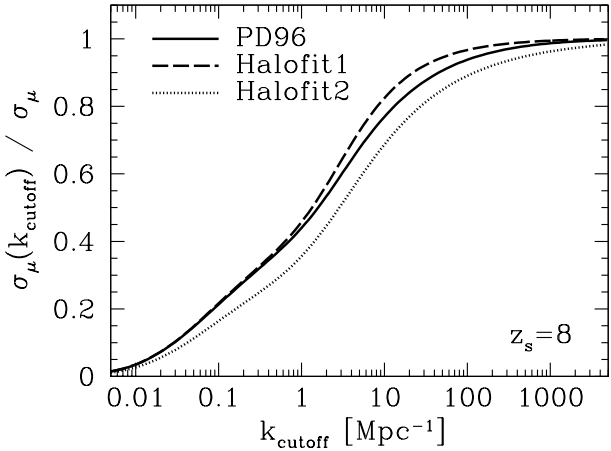}}
\end{center}
\caption{Ratio of the standard deviation of $\mu$, assuming a cutoff wavenumber in the matter power spectrum (above which the power spectrum is set to zero), to the full standard deviation with no cutoff.  Here, we consider the case of a blank field, and assume a fixed source redshift of $\zs=8$.  The standard deviation of $\mu$ receives contribution from power spanning a wide range of wavenumbers, from $k\sim5\times10^{-3}\Mpc^{-1}$ up to $k\sim 1000 \Mpc^{-1}$. }
\label{FIG:kcutoff}
 \end{figure}

In this section, we explore the $k$-range of the power spectrum that contributes significantly to the magnification dispersion.  For this purpose, we focus on the blank field result, and introduce a cutoff wavenumber $\kcutoff$, above which the power spectrum $P_{\Phi}$ is set to zero in equation (\ref{EQ:covariance}).  We then calculate the standard deviation of $\mu$ as a function of $\kcutoff$ and divide by the full calculation in which there is no cutoff.  Figure \ref{FIG:kcutoff} shows the result of this exercise for the PD96, Halofit1, and Halofit2 power spectra.  These results are nearly independent of the source redshift for $\zs \gtrsim 6$, so we show only the $\zs=8$ case in the figure.   The standard deviation of $\mu$ receives contribution from power over a wide range of scales, starting in the linear regime at $k\sim5\times10^{-3}~\Mpc^{-1}$, and does not converge until after $k\sim1000~\Mpc^{-1}$. Note that the rate of convergence to the full result depends on the amount of small-scale power.  For example, the slowest to converge is Halofit2, which exhibits the most small-scale power among the three models (see Figure \ref{FIG:powerspectra}).

A more detailed extension of the work presented in this paper will likely require numerical techniques, such as ray tracing through simulations of cosmological structure formation.  The wide range of scales that contribute to the magnification dispersion, from linear to highly nonlinear, illustrates the importance of using large-scale, high-resolution simulations for studying CWL magnification \citep[see also][]{2011ApJ...742...15T}.  The calculations presented here provide a convenient way to estimate the errors resulting from finite numerical resolution, insofar as $\kcutoff$ crudely mimics the smallest scale resolved by a simulation.  For example, if the smallest scale resolved roughly corresponds to a wavenumber of $k\sim10~\Mpc^{-1}$ (corresponding to a mass-scale of $M\sim\bar{\rho}~4 \pi/3~(2 \pi/k)^{3}  = 4\times10^{10}\Msun$), the dispersion of the CWL magnification could be underestimated by as much as $\sim20~\%$, according to Figure \ref{FIG:kcutoff}.                  

\section{Summary and Discussion}
\label{SEC:Conclusion}

In cluster-lensing systems, light rays from background galaxies are lensed not just by the cluster, but also by intervening LSS between the source and observer.  Since lens reconstruction methodologies almost always ignore the effect of this CWL, it could be a source of error in the magnification maps of cluster-lenses, particularly for high-$z$ sources.  Here, we have assessed the contribution of CWL to the magnifications of high-$z$ galaxies observed in cluster-lensing systems.   

We first quantified the level of typical magnification fluctuations in a blank field (no foreground cluster, with $\langle \mu \rangle=1$) by calculating the standard deviation of $\mu$.  We found typical (1$\sigma$) fluctuations in $\mu$ to be $\sim17-24\%$ for galaxies at $\zs=6$, rising modestly to $\sim21-28\%$ by $\zs=12$.  We then used a simple model, in which cluster mass profiles take the NFW form, to quantify statistically the effect of CWL in magnification maps of cluster-lenses.  To measure this effect, we calculated the fractional standard deviation of $1/\mu$ as a function of angular displacement from lens-center.  For lenses at $\zL=0.3-0.5$, and sources at $\zs=6-10$, we found this fractional standard deviation to range from $\sim 10-20\%$ for $|\mu| \lesssim 5$, to $\sim 20-30\%$ for $|\mu| \sim 10$. However, the fractional standard deviation rises to greater than order unity near critical curves, implying that CWL may be an important consideration for measuring the intrinsic luminosities of the most magnified galaxies. This phenomenon originates from the fact that CWL perturbs the critical curves of the primary lens, placing them either closer to or further away from a nearby test image, depending on the particular line-of-sight.  

Future work should extend these results by assessing the contribution of CWL numerically with more realistic models of cluster-lenses, beyond the smooth and symmetric model adopted here.  Our semi-analytical calculations inform such future efforts.   For example, we found that the magnification dispersion in a blank field does not fully converge until after $k\sim 1000 \Mpc^{-1}$, indicating that our results are sensitive to the matter power spectrum at large wavenumber.   On the other hand, we also found that the dispersion receives contribution from large-scale power, corresponding to wavenumbers as low as $k\sim5\times10^{-3}\Mpc^{-1}$.  These findings highlight the need to employ cosmological simulations that are simultaneously large-enough in volume, and high-enough in resolution, to capture the full range of structures contributing to CWL magnification.  

In this paper, we showed that it can be important to account for CWL in lens models, especially for clusters with highly-magnified images\footnote{After the submission of this manuscript, a paper by \citet{2014ApJ...783...41B} appeared on the arXiv preprint archive.  Using cluster-lenses from the Sloan Giant Arcs Survey and spectroscopically identified foreground and background groups in the fields of those clusters, \citet{2014ApJ...783...41B}  reported evidence that their strong-lensing sample is biased towards lines-of-sight with an excess of uncorrelated LSS.  If observed cluster-lenses tend to lie along such lines of sight, i.e. those in which CWL contributes more than average to the lensing cross-section and magnification, then further study of the impact of CWL on magnification measurements is important for current/future surveys aiming to exploit cluster-lenses as cosmic telescopes.}.  However, we have not attempted to assess the magnification measurement error which results from constructing a lens-model that neglects CWL.  That is beyond the scope of this paper.   There is considerable variation among existing lens reconstruction methodologies \citep[see e.g.][and references therein]{2011A&ARv..19...47K}, and the magnification measurement error likely varies significantly from model to model.  In the ongoing \emph{HST} \emph{Frontier Fields} model comparison project, it would be useful to generate mock data from simulations which combine a simple, idealized cluster-lens with CWL, to test how well existing lens models can recover the ``true" magnifications of high-$z$ galaxies by neglecting CWL\footnote{Since the submission of this manuscript, the \emph{Frontier Fields} model comparison project has commenced work on this test.  The publication of their results is forthcoming.}.    
 
Finally, we emphasize that the results presented here \emph{do not} preclude accurate intrinsic luminosity measurements for highly-magnified galaxies.  In fact, it may be possible to account for the effects of CWL on these images through careful modeling.  For example, in the case of galaxy-strong-lensing, which typically involves only one strongly-lensed background source, it has been shown that CWL can in principle be fully modeled by an external convergence and shear, plus adjustment of the primary lens-potential ellipticity \citep{1996ApJ...468...17B,1997ApJ...482..604K,1997MNRAS.292..673S}.  Now that we have demonstrated the importance of modeling CWL for highly-magnified images, it would be useful to extend those analyses to the case of cluster-lensing, in which lenses exhibit images of many background sources at different redshifts.  Prescriptions to account for the contribution of CWL to image locations in cluster-strong-lensing error analyses have been presented in \citet{Jullo2010Sci} \citep[see also][]{2011MNRAS.411.1628D} and \citet{2012MNRAS.420L..18H}.  It would also be useful to extend those error analysis methods to account for the contribution of CWL to relative image fluxes.       

In the next few years, cluster-lenses acting as gravitational telescopes will provide a first view of high-$z$ sources that would otherwise be inaccessible until the advent of next-generation telescopes, and may even provide a first measurement of the luminosity function of such sources.  Our results motivate new approaches to improving magnification measurements with gravitational telescopes.

\section*{Acknowledgments}
A.D. thanks Eiichiro Komatsu and Steven Finkelstein for useful discussions.  We thank the anonymous referee for helpful comments. This work was supported in part by U.S. NSF grants AST-0708176 and AST-1009799, and NASA grants NNX07AH09G and NNX11AE09G.  P.N. gratefully acknowledges support from an NSF theory program via the grant AST-1044455 and a theory grant from the Space Telescope Science Institute HST-AR-12144.01-A.

\bibliographystyle{mn2e}
\bibliography{master}

\end{document}